\documentclass[runningheads]{llncs}
\usepackage[T1]{fontenc}
\usepackage[utf8]{inputenc}
\usepackage{graphicx}
\usepackage{amsmath}
\usepackage{amssymb}
\usepackage{mathtools}
\usepackage{booktabs}
\usepackage{array}
\usepackage{multirow}
\usepackage{algorithm}
\usepackage[noend]{algpseudocode}
\usepackage{url}
\usepackage{xcolor}
\usepackage{hyperref}

\algnewcommand\algorithmicforeach{\textbf{foreach}}
\algdef{S}[FOR]{ForEach}[1]{\algorithmicforeach\ #1\ \algorithmicdo}
\algrenewcommand\algorithmicindent{.5em}

\renewcommand{\arraystretch}{1.125}

\begin{document}
\newcolumntype{L}[1]{>{\raggedright\arraybackslash}p{#1}}

\title{Leakage-Safe Empirical Benchmarking of EEG-Based Machine Learning Pipelines for Dementia Classification}
\titlerunning{Leakage-Safe EEG-ML Benchmark for Dementia Classification}

\author{
Haitian Wang\inst{1}
\and Chamara Madarasingha\inst{1}\thanks{Corresponding author: \email{C.Kattadige@curtin.edu.au}}
\and Redowan Mahmud\inst{1}
\and \\ Aneesh Krishna\inst{1}
\and Ryu Takechi\inst{2}
}

\authorrunning{H. Wang et al.}

\institute{
School of Electrical Engineering, Computing and Mathematical Sciences, Curtin University, bentley, WA, Australia
\and
School of Population Health, Curtin University, Bentley, WA, Australia
}

\maketitle

\vspace{-7mm}
\begin{abstract}
Electroencephalography (EEG) is a low-cost and non-invasive signal source for dementia screening, yet existing EEG-based studies remain difficult to compare because preprocessing, EEG segmentation, feature design, classifier choice, and validation protocols vary across studies and are often evaluated in isolation. This variability limits the derivation of robust pipeline recommendations. This paper presents a leakage-safe empirical benchmark for resting-state EEG-based dementia classification and uses it to identify a practical best-practice pipeline. Using the public OpenNeuro ds004504 dataset, the benchmark evaluates artifact correction, fixed-length EEG segmentation, training-only augmentation, multi-domain feature extraction, fold-internal feature selection, classical machine-learning (ML) classifiers, subject-level aggregation, and interpretation under leave-one-subject-out (LOSO) validation. The best-performing pipeline in this benchmark combines Artifact Subspace Reconstruction (ASR) followed by Independent Component Analysis (ICA), 10 s EEG epochs with training-only amplitude scaling and Gaussian-noise augmentation, spectral, complexity, and pairwise connectivity features, mutual-information top-100 selection, linear support vector machine classification, and mean-probability aggregation. It achieves 87.69\% accuracy, 88.89\% F1-score, and 91.20\% AUC on Alzheimer's disease (AD) versus cognitively normal controls (CN) classification while retaining 100 features from 1596 raw descriptors. The results support compact multi-domain EEG features as an interpretable and leakage-safe baseline for subject-level dementia classification. To support reproducibility, the implementation of the leakage-safe EEG-ML benchmark, including preprocessing, feature extraction, feature selection, model evaluation, and LOSO validation scripts, is released at \href{https://github.com/HaitianWang/Leakage-Safe-EEG-ML-Benchmark-for-Dementia-Classification}{\textcolor{blue}{GitHub Repository}}.
\vspace{-7mm}
\end{abstract}

%-------------------------------------------------------------------------------
\vspace{-8mm}
\section{Introduction}
\vspace{-3mm}

Dementia is a major neurological condition characterised by progressive decline in memory, cognition, and daily functioning. Alzheimer's disease (AD) is the most common form of dementia, while frontotemporal dementia (FTD) can present with overlapping symptoms but distinct pathological mechanisms~\cite{scheltens2021alzheimers,antonioni2023frontotemporal}. Although MRI, fMRI, and PET provide valuable information for dementia diagnosis, their cost, limited accessibility, and acquisition constraints restrict their use as scalable screening tools. Electroencephalography (EEG) offers a low-cost, non-invasive, and high-temporal-resolution alternative for measuring brain activity, making it suitable for machine-learning (ML)-based clinical-assistive screening rather than replacing clinical diagnosis~\cite{bi2025eegreview,akbar2025unlocking}. Existing EEG-based dementia studies, including calibrated subject-wise models, microstate-based features, and deep feature-fusion approaches, have reported promising results~\cite{zheng2023diagnosis,rostamikia2024eeg,jiang2025coherence,shamsi2025reliable,yang2024microstate,akbar2025neurofusionnet}; however, their findings remain difficult to compare and translate because they use heterogeneous preprocessing pipelines, EEG-segment lengths, feature representations, classifiers, and validation protocols. Feature designs range from spectral power descriptors~\cite{zheng2023diagnosis,wang2025rhythmic,zandbagleh2024entropy} to graph-based, connectivity-based, and learned representations~\cite{si2023differentiating,ma2024classification,jiang2025coherence,mlinari2026functional,miltiadous2023dice}, but most studies evaluate only selected components rather than the complete EEG-ML workflow. Moreover, DL models are data-demanding, small EEG cohorts increase the risk of subject-specific overfitting and reduced interpretability, and unclear subject-level separation can introduce EEG-segment-level data leakage that inflates performance. Therefore, a key methodological gap remains: the field lacks a controlled, reproducible, and leakage-safe benchmark that systematically evaluates preprocessing, feature engineering, dimensionality control, classifier selection, and subject-level aggregation under the same subject-level validation protocol.

% \begin{figure}[t]
%     \centering
%     \includegraphics[width=1.0\textwidth]{eeg_dementia_pipeline_intro.png}
%     \vspace{-8mm}
%     \caption{Pipeline of the proposed leakage-safe EEG-ML benchmark for dementia classification.}
%     \label{fig:pipeline_intro}
%     \vspace{-8mm}
% \end{figure}

To address this gap, we propose a leakage-safe empirical benchmark for EEG-based dementia classification and instantiate it through a reproducible classical ML pipeline. Using the public OpenNeuro ds004504 dataset with AD, FTD, and cognitively normal subjects, the benchmark defines the complete workflow from raw resting-state EEG to subject-level prediction, including preprocessing, fixed-length EEG segmentation, training-only augmentation, multi-domain feature extraction, feature selection, classification, probability-based aggregation, evaluation, and interpretation~\cite{miltiadous2023dataset}. On this basis, we conduct a controlled empirical evaluation of the main pipeline choices, including filtering, Independent Component Analysis (ICA), Artifact Subspace Reconstruction (ASR), and ASR followed by ICA (ASR + ICA) for artifact correction, spectral, complexity, pairwise-connectivity, and connectivity-plus-graph feature settings, fold-internal dimensionality control, and representative classical classifiers such as LDA, Logistic Regression, KNN, Random Forest, ExtraTrees, and Gradient Boosting. All augmentation, feature selection, model selection, and performance reporting are performed under strict subject-level validation, allowing the benchmark to quantify the effects of each component on accuracy, compactness, interpretability, and computational cost while avoiding EEG-segment-level leakage. 

% Fig.~\ref{fig:pipeline_intro} summarises the evaluated EEG-ML pipeline from resting-state EEG input to subject-level aggregation and evaluation.

The staged benchmark identifies the best-performing subject-level pipeline in this study, consisting of ASR + ICA artifact correction, 10 s EEG epochs with training-only amplitude scaling and Gaussian-noise augmentation, spectral, complexity, and pairwise connectivity features, mutual-information top-100 feature selection, linear SVM classification, and mean-probability aggregation. On AD versus CN classification, this configuration achieves 87.69\% accuracy, 88.89\% F1-score, and 91.20\% AUC, outperforming the classical counterpart baselines evaluated under the same subject-level leave-one-subject-out (LOSO) protocol. The same pipeline achieves 82.69\% accuracy for FTD versus CN, 77.97\% for AD versus FTD, and 70.45\% for the three-class task. It retains 100 features from 1596 raw descriptors, and the feature-level analysis shows complementary contributions from spectral, complexity, and connectivity descriptors, with prominent temporal-parietal and frontal-temporal patterns.

The main contributions of this paper are as follows:
\vspace{-2mm}
\begin{itemize}
    \item We introduce a leakage-safe empirical benchmark for EEG-based dementia classification, covering raw EEG preparation, preprocessing, fixed-length EEG segmentation, training-only augmentation, multi-domain feature engineering, feature selection, classical ML classification, subject-level aggregation, evaluation, and feature-level analysis.
    \item We instantiate the benchmark through a reproducible classical ML pipeline on the widely used OpenNeuro ds004504 benchmark dataset, with all augmentation, feature selection, model selection, and performance reporting performed under subject-level separation.
    \item We conduct controlled empirical experiments to benchmark preprocessing strategies, feature families, dimensionality control methods, and classical ML classifiers under leave-one-subject-out validation.
    \item We provide a practical best-practice recommendation for EEG-based dementia classification by identifying pipeline configuration that balances subject-level performance, feature dimensionality, computational cost, and region-level interpretability.
\end{itemize}

%-------------------------------------------------------------------------------
\vspace{-5mm}
\section{Related Work}
\label{sec:RelatedWork}
\vspace{-3mm}

Resting-state EEG has been widely used for dementia analysis because it is non-invasive, low-cost, and sensitive to changes in neural rhythms, signal complexity, and inter-channel synchronisation. Prior reviews and recent methodological studies have reported dementia-related EEG patterns such as spectral slowing, reduced nonlinear complexity, and altered functional connectivity, supporting EEG as a clinical-assistive signal source for Alzheimer's disease and related dementias~\cite{bi2025eegreview,akbar2025unlocking,ghassemkhani2025complexity}. The widely used OpenNeuro ds004504 benchmark dataset further supports reproducible research by providing resting-state EEG recordings from Alzheimer's disease, frontotemporal dementia, and cognitively normal subjects under a shared acquisition protocol~\cite{miltiadous2023dataset}. Recent studies on this dataset have further emphasised subject-wise evaluation, calibrated probabilities, and alternative EEG feature families such as wavelet-scattering statistics and microstate descriptors~\cite{shamsi2025reliable,yang2024microstate}. These works strengthen the need for leakage-free evaluation, but they mainly focus on a specific feature representation or probability-calibration pipeline rather than a staged comparison of preprocessing, feature families, dimensionality control, and classical classifiers.

AI-based EEG dementia studies follow feature-based ML or representation-learning strategies. Classical ML methods use spectral power, nonlinear complexity, microstate, connectivity, and graph-related descriptors to construct compact biomarkers, which is useful for small EEG cohorts and for analysing frequency-band, channel, and region-level effects~\cite{zheng2023diagnosis,vicchietti2023computational,yang2024microstate,rostamikia2024eeg,ma2024classification,senkaya2025enhancing}. DL and feature-fusion methods, including CNN, transformer, graph-based, and hybrid architectures, learn task-specific representations from EEG features or transformed EEG inputs~\cite{miltiadous2023dice,jiang2025coherence,stefanou2025cnn,umair2025privacy,akbar2025neurofusionnet}. NeuroFusionNet is a recent example that combines handcrafted EEG features with CNN-derived latent embeddings and explainability analysis, but its emphasis is hybrid deep feature fusion rather than isolating the contribution of each classical EEG-ML pipeline component~\cite{akbar2025neurofusionnet}. 

Although these studies provide useful model designs and performance baselines, most focus on a selected feature family, a specific classifier, or a single task. This variability in design choices makes it difficult to determine whether performance differences are caused by signal cleaning, feature engineering, dimensionality control, model capacity, or validation design, thereby limiting their practical value for deriving robust pipeline recommendations. This issue is critical for small EEG datasets, where high-dimensional features and EEG-segment-level samples can increase overfitting and inflate performance if all EEG epochs from the same subject are not kept within the same validation fold.

\vspace{-4mm}
\section{Dataset and Materials}
\label{sec:dataset}
\vspace{-3mm}

We use the public OpenNeuro ds004504 benchmark dataset~\cite{miltiadous2023dataset}, which contains resting-state eyes-closed EEG recordings from subjects with AD, FTD and CN. The dataset was selected because it provides all three target groups in a single acquisition protocol, includes both raw and preprocessed BIDS-compatible EEG files~\cite{miltiadous2023dataset}, and has been reused in recent EEG-based dementia classification studies. In this work, the raw recordings are used as the starting point so that preprocessing choices can be controlled within the benchmark.

\begin{table}[t]
\centering
\caption{Summary of the OpenNeuro ds004504 dataset used in this study.}
\vspace{-3mm}
\label{tab:dataset_summary}
\scriptsize
\begin{tabular}{p{0.16\linewidth}p{0.16\linewidth}p{0.18\linewidth}p{0.34\linewidth}}
\toprule
Group & Subjects & Mean duration & Role in benchmark \\
\midrule
AD & 36 & 13.5 min & Alzheimer's disease group for disease-control and dementia-subtype classification \\
FTD & 23 & 12.0 min & Frontotemporal dementia group for differential dementia classification \\
CN & 29 & 13.8 min & Cognitively normal control group for disease-control classification \\
\midrule
Total & 88 & 13.2 min & Three-class benchmark with subject-level validation \\
\bottomrule
\end{tabular}
\vspace{-6mm}
\end{table}

Each subject was recorded using a clinical EEG system with 19 scalp electrodes placed according to the international 10--20 system~\cite{miltiadous2023dataset}. The released channels are Fp1, Fp2, F7, F3, Fz, F4, F8, T3, C3, Cz, C4, T4, T5, P3, Pz, P4, T6, O1, and O2. The original sampling rate is 500 Hz, and each subject is associated with a Mini-Mental State Examination score~\cite{miltiadous2023dataset}. The recordings were acquired using a monopolar montage, and the dataset includes EEG data suitable for analysing spectral activity, channel-level signal dynamics, and inter-channel connectivity.

\begin{figure}[t]
    \centering
    \includegraphics[width=0.8\linewidth]{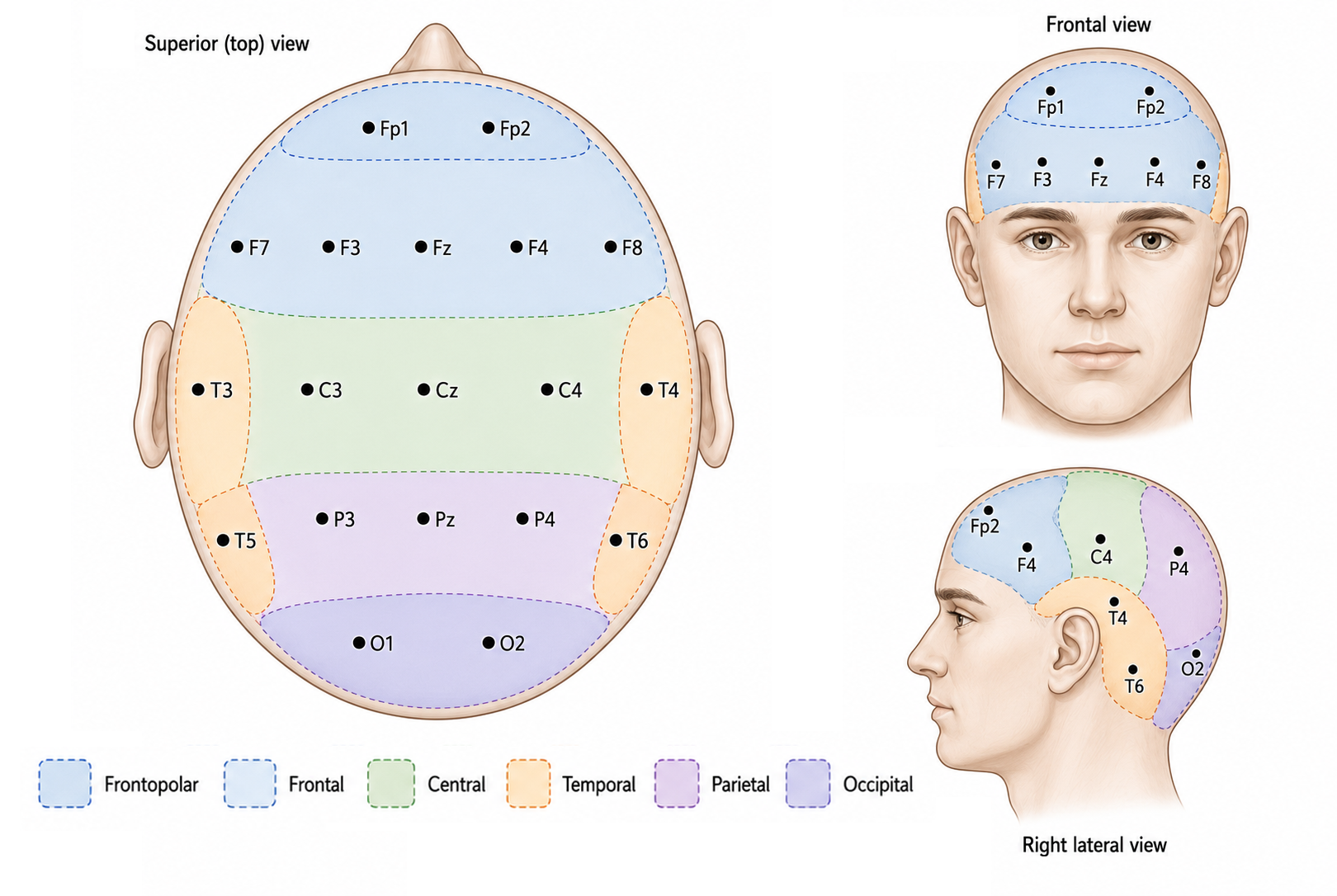}
    \vspace{-8mm}
    \caption{Spatial layout of the 19 EEG channels used in the benchmark. Frontopolar and frontal electrodes are visually separated for anatomical clarity, but they are merged as the frontal region in region-level feature aggregation.}
    \label{fig:eeg_channels}
    \vspace{-6mm}
\end{figure}

For region-level analysis, the 19 channels are grouped into frontal, temporal, central, parietal, and occipital regions. The frontal region contains Fp1, Fp2, F7, F3, Fz, F4, and F8. The temporal region contains T3, T4, T5, and T6. The central region contains C3, Cz, and C4. The parietal region contains P3, Pz, and P4. The occipital region contains O1 and O2. This grouping is used for feature aggregation and interpretation. Model training remains subject-independent and all EEG epochs from the same subject are kept within the same fold.

\vspace{-3mm}
\section{Methodology}
\label{sec:Methodology}
\vspace{-2mm}

\begin{figure}[t]
    \centering
    \includegraphics[width=\textwidth]{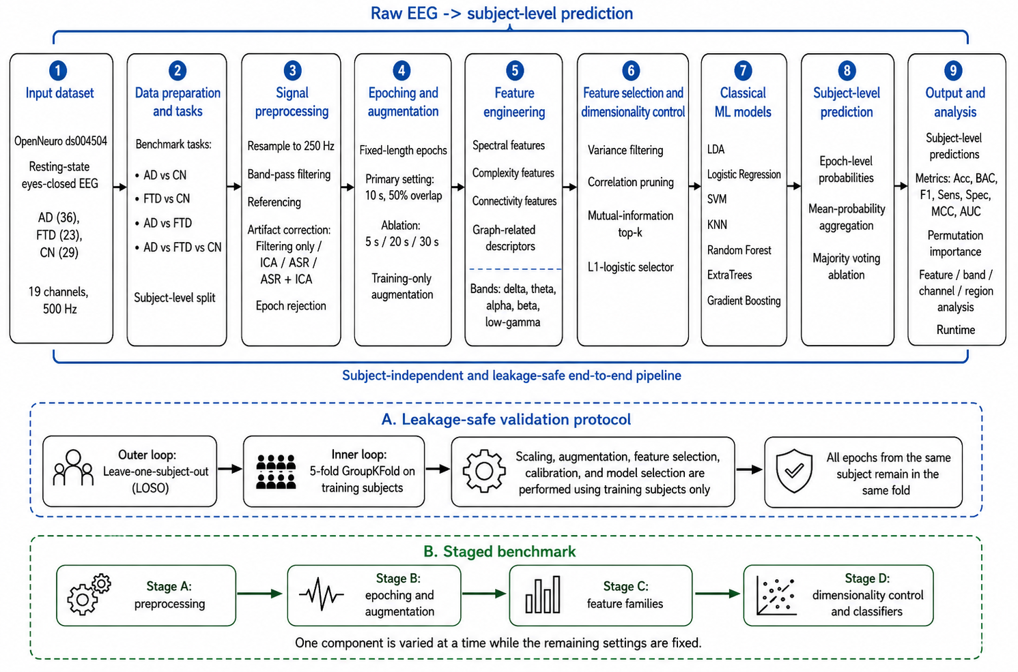}
    \vspace{-6mm}
    \caption{End-to-end leakage-safe EEG-ML benchmark pipeline for dementia classification. The pipeline converts raw resting-state EEG from the OpenNeuro ds004504 dataset into subject-level predictions through data preparation, signal preprocessing, fixed-length EEG segmentation and training-only augmentation, feature engineering, feature selection, classical ML classification, and subject-level analysis. The lower panels summarise the leakage-safe validation protocol and the staged benchmark used to evaluate preprocessing, EEG-segment length, feature families, dimensionality control, and classifier selection.}
    \label{fig:pipeline}
    \vspace{-6mm}
\end{figure}

The benchmark pipeline converts raw EEG recordings into subject-level predictions through data preparation, signal preprocessing, fixed-length EEG segmentation, training-only augmentation, feature engineering, feature selection, ML classification, probability aggregation, evaluation, and interpretation, as shown in Fig.~\ref{fig:pipeline}.

\vspace{-3mm}
\subsection{Data Preparation, EEG Segmentation, Augmentation, and Subject-Level Partitioning}
\vspace{-2mm}

This subsection defines the task setup, EEG segmentation, training-only augmentation, and subject-level partitioning used before feature extraction and classification.

\subsubsection{Data Preparation and Benchmark Tasks}
\vspace{-2mm}

Let $\mathcal{D}=\{(X_i,y_i,s_i)\}_{i=1}^{N}$ denote the subject-level dataset described in Section~\ref{sec:dataset}, where $X_i \in \mathbb{R}^{C \times T_i}$ is the EEG recording of subject $i$, $C$ is the number of EEG channels, $T_i$ is the number of time samples, $y_i \in \{\mathrm{AD},\mathrm{FTD},\mathrm{CN}\}$ is the diagnosis label, and $s_i$ is the subject identifier.

Four benchmark tasks are evaluated: AD versus CN, FTD versus CN, AD versus FTD, and AD versus FTD versus CN. The binary tasks measure disease-control and dementia-subtype separability, while the three-class task evaluates differential screening among two dementia groups and controls. The primary validation protocol is leave-one-subject-out cross-validation. In each outer fold, all EEG epochs from one subject are held out for testing, and scaling, augmentation, feature selection, model selection, calibration, and training are performed only on the remaining subjects. The inner loop uses 5-fold GroupKFold with subject identifiers as groups.

\vspace{-3mm}
\subsubsection{EEG Segmentation and Training-Only Augmentation}
\vspace{-2mm}

After preprocessing, each recording is divided into fixed-length EEG epochs, where an EEG epoch refers to a time-windowed signal segment rather than one full pass of ML model training. The primary setting uses 10 s EEG epochs with 50\% overlap. This duration retains enough cycles for delta and theta estimation while providing more training samples than 20 s or 30 s windows. EEG-epoch lengths of 5 s, 20 s, and 30 s are used only in the segmentation-length ablation. For subject $i$, the epoch set is defined as
\begin{small}
\begin{equation}
\mathcal{E}_i =
\left\{
E_{i,j}=X_i[:,\,jh,\ldots,jh+L-1]
\,\middle|\,
j=0,\ldots,\left\lfloor\frac{T_i-L}{h}\right\rfloor
\right\},
\quad
L=f_s\tau,
\quad
h=\frac{L}{2}.
\label{eq:epoching}
\end{equation}
\end{small}
Here, $E_{i,j}\in\mathbb{R}^{C\times L}$ is the $j$-th EEG epoch of subject $i$, the colon denotes all EEG channels, $f_s$ is the resampled frequency, $\tau$ is the epoch duration, $L$ is the number of samples per epoch, and $h$ is the stride.

To reduce overfitting caused by the limited number of subjects, a training-only augmentation setting is evaluated as a controlled ablation. For each EEG epoch from the training subjects, one additional training instance is derived by channel-wise amplitude scaling and low-level Gaussian noise:
\begin{small}
\begin{equation}
\widetilde{E}_{i,j,c}(t)
=
a_cE_{i,j,c}(t)+\epsilon_{c}(t),
\quad
a_c \sim \mathcal{U}(0.95,1.05),
\quad
\epsilon_c(t)\sim\mathcal{N}\left(0,(0.01\sigma_c)^2\right).
\label{eq:augmentation}
\end{equation}
\end{small}
In Eq.~\ref{eq:augmentation}, $c$ indexes an EEG channel, $t$ indexes a time sample within the EEG epoch, and $\sigma_c$ is the standard deviation of channel $c$ within the original training EEG epoch. This augmentation is applied only inside the training fold before feature extraction. Validation and test subjects are never modified. The effects of amplitude scaling and Gaussian noise are evaluated in the staged ablation study.

\vspace{-3mm}
\subsubsection{Subject-Level Aggregation}
\vspace{-2mm}

The classifier is trained on EEG-segment-level samples, while predictions are aggregated at the subject level. For a held-out subject, EEG-segment-level class probabilities are averaged across all valid EEG epochs. The predicted subject label is the class $q$ with the largest mean probability:
\vspace{-2mm}
\begin{small}
\begin{equation}
\widehat{y}_i
=
\operatorname*{arg\,max}_{q \in \mathcal{Y}}
\frac{1}{|\mathcal{E}_i|}
\sum_{E_{i,j}\in\mathcal{E}_i}
p_{\theta}(y=q \mid E_{i,j}).
\label{eq:aggregation}
\end{equation}
\vspace{-2mm}
\end{small}

Mean probability aggregation is used as the subject-level rule because it retains classifier confidence and prevents EEG epochs from being treated as independent clinical cases.

\vspace{-3mm}
\subsection{Signal Preprocessing}
\vspace{-2mm}

This subsection specifies the preprocessing stage used to standardise raw EEG recordings before EEG segmentation and feature extraction. It defines filtering, referencing, artifact correction, and EEG-epoch rejection, and provides the preprocessing variants compared in the staged benchmark.

\vspace{-3mm}
\subsubsection{Filtering and Referencing}
\vspace{-2mm}

Raw EEG files are loaded in BIDS-compatible format and mapped to the 10--20 montage following the dataset specification~\cite{miltiadous2023dataset}. All recordings are resampled from 500 Hz to 250 Hz. This sampling rate preserves the analysed 0.5--45 Hz range and reduces computational cost for feature extraction. A zero-phase finite impulse response band-pass filter from 0.5 Hz to 45 Hz is applied to suppress slow drift and high-frequency muscular contamination, following common EEG preprocessing practice in recent dementia studies~\cite{bi2025eegreview,akbar2025unlocking}. This range is intentionally wider than the 0.5--40 Hz or 1--40 Hz settings often used in EEG dementia pipelines, allowing the benchmark to test whether retaining low-gamma activity contributes to classification after artifact correction. Since the upper cutoff is below 50 Hz, a separate 50 Hz notch filter is not applied in the reference pipeline.

After artifact correction, the signal is re-referenced to the common average reference. Channels detected as corrupted before re-referencing are reconstructed by spherical interpolation when at most two channels are affected. Recordings with more than two unusable channels are excluded from the corresponding fold to avoid unstable connectivity estimates.

\vspace{-3mm}
\subsubsection{Artifact Correction}
\vspace{-2mm}

EEG recordings contain physiological and non-physiological artifacts such as eye movement, blinking, muscle activity, electrode instability, and transient high-amplitude bursts. The benchmark evaluates four artifact correction pipelines: filtering only, ICA-only, ASR-only, and ASR followed by ICA. The main selected pipeline uses ASR followed by ICA because ASR attenuates burst-like artifacts before ICA separates residual ocular and muscular components. ICA-only is retained as a reproducible ablation because it directly targets ocular and muscular components without requiring manual epoch removal.

ICA is fitted on the filtered continuous signal with the number of components equal to the data rank, a maximum of 1000 iterations, and a fixed random seed. Components labelled as eye, muscle, heart, line noise, or channel noise with probability greater than 0.80 by ICLabel~\cite{pion2019iclabel} are removed. ASR is implemented with the clean\_rawdata procedure in the ASR ablation. ICA- and ASR-based cleaning strategies are commonly used in recent EEG-based dementia studies to attenuate ocular, muscular, channel, and burst-like artifacts~\cite{bi2025eegreview,rostamikia2024eeg}. The burst cutoff is set to 20 standard deviations, the flatline threshold to 5 s, the channel correlation threshold to 0.80, and the maximum bad-window channel proportion to 0.25. ASR is not used to tune labels or model parameters.

\vspace{-3mm}
\subsubsection{EEG-Epoch Rejection}
\vspace{-2mm}

EEG epochs are rejected after segmentation if any channel has peak-to-peak amplitude above 150 $\mu$V or if the EEG-epoch log-variance deviates by more than 3.5 standard deviations from the subject-level median. This automatic criterion removes residual bursts that remain after filtering and source correction. It is applied before feature extraction and independently within each subject.

\vspace{-3mm}
\subsection{Feature Engineering}
\vspace{-2mm}

All classifiers receive tabular EEG feature vectors rather than raw waveforms, spectrogram images, or scalograms. This design matches the classical ML focus of the study and allows feature, channel, band, and region-level interpretation. Features are computed for five frequency bands: delta 0.5-4 Hz, theta 4-8 Hz, alpha 8-13 Hz, beta 13-30 Hz, and low-gamma 30-45 Hz. The low-gamma band is retained for comparison, but interpretation is centred on delta, theta, alpha, and beta because scalp low-gamma is more sensitive to muscle artifacts.

\vspace{-3mm}
\subsubsection{Spectral Features}
\vspace{-2mm}

Spectral features quantify EEG slowing and band-specific power redistribution. PSD is estimated by Welch's method~\cite{nardone2024comparison} using a Hann window of 4 s, 50\% window overlap, and $n_{\mathrm{fft}}=1024$, following recent EEG-based dementia studies that use spectral power and PSD-derived descriptors for AD/FTD classification~\cite{zheng2023diagnosis,senkaya2025enhancing,ghassemkhani2025complexity}.
For channel $c$, the PSD estimate is
\begin{small}
\begin{equation}
\widehat{S}_{cc}(f)=
\frac{1}{K f_s U}
\sum_{k=0}^{K-1}
\left|
\sum_{n=0}^{M-1}
w[n]x_c[n+kR]
\exp\left(-\mathrm{i}2\pi fn/f_s\right)
\right|^2,
\quad
U=\frac{1}{M}\sum_{n=0}^{M-1}w^2[n].
\label{eq:welch}
\end{equation}
\end{small}
Here, $x_c$ is the signal from channel $c$, $M$ is the window length, $R$ is the window stride, $K$ is the number of windows, $w[n]$ is the Hann window, and $U$ is the window normalisation term. For band $b=[f_1,f_2]$, band power is computed as $P_c^{(b)}=\int_{f_1}^{f_2}\widehat{S}_{cc}(f)\,df$. For each channel, we extract a compact 14-dimensional spectral descriptor set, including band-power features, spectral entropy, alpha peak frequency, and clinically motivated slow-to-fast rhythm ratios. These descriptors test whether dementia groups show increased slow activity relative to faster rhythms.

\vspace{-3mm}
\subsubsection{Complexity Features}
\vspace{-2mm}

Complexity features describe the temporal irregularity and waveform structure of each channel. They are extracted as a compact 25-dimensional descriptor set per channel from broadband and selected band-limited EEG epochs. Hjorth activity, mobility, and complexity are computed as compact time-domain descriptors commonly used in EEG-based dementia feature engineering~\cite{zheng2023diagnosis,bi2025eegreview}. Activity is the variance of the signal, mobility is $\sqrt{\mathrm{Var}(\dot{x})/\mathrm{Var}(x)}$, and Hjorth complexity is the ratio between the mobility of $\dot{x}$ and the mobility of $x$.

Sample entropy is computed with embedding dimension $m=2$ and tolerance $r=0.2\sigma$, where $\sigma$ is the standard deviation of the analysed EEG epoch, to quantify nonlinear signal irregularity in a compact EEG complexity representation~\cite{zheng2023diagnosis,ghassemkhani2025complexity}. For time series $x$, let $B_m(r)$ denote the number of matched template pairs of length $m$ and $B_{m+1}(r)$ the number of matched template pairs of length $m+1$. The feature is
\begin{small}
\begin{equation}
\mathrm{SampEn}(m,r,N)
=
-\log
\frac{B_{m+1}(r)}
{B_m(r)}.
\label{eq:sampen}
\end{equation}
\end{small}
Higuchi fractal dimension is computed with $k_{\max}=8$ to quantify the geometric irregularity of each EEG segment. A compact multi-scale entropy feature is also included with scales $\{1,2,3,4,5\}$ using the same $m$ and $r$ parameters. These complexity descriptors are evaluated because recent EEG dementia studies show that nonlinear and fractal features can capture disease-related changes beyond simple power redistribution~\cite{vicchietti2023computational,ghassemkhani2025complexity,bi2025eegreview}.

\vspace{-3mm}
\subsubsection{Connectivity and Graph Features}
\vspace{-2mm}

Connectivity features quantify functional coordination between scalp channels. With 19 channels, each EEG epoch contains 171 undirected channel pairs. We compute the weighted phase lag index (wPLI) for each channel pair and frequency band, following recent EEG dementia studies that use phase-based and functional-connectivity descriptors for AD/FTD discrimination~\cite{si2023differentiating,mlinari2026functional}. This yields $171 \times 5 = 855$ pairwise connectivity descriptors per EEG epoch before feature selection. For channels $c$ and $d$, the wPLI feature is
\begin{small}
\begin{equation}
\mathrm{wPLI}_{cd}^{(b)}
=
\frac{
\left|
\mathbb{E}
\left[
\left|\Im(Z_{cd}^{(b)})\right|
\mathrm{sign}\left(\Im(Z_{cd}^{(b)})\right)
\right]
\right|
}{
\mathbb{E}\left[\left|\Im(Z_{cd}^{(b)})\right|\right]+\epsilon
}.
\label{eq:connectivity}
\end{equation}
\end{small}
Here, $Z_{cd}^{(b)}$ is the band-limited cross-spectrum between channels $c$ and $d$, $\Im(\cdot)$ denotes the imaginary component, and $\epsilon=10^{-8}$ prevents division by zero. wPLI reduces the influence of zero-lag synchronisation caused by volume conduction.

Pairwise connectivity features are aggregated into five anatomical regions: frontal, temporal, central, parietal, and occipital. The frontal region contains Fp1, Fp2, F7, F3, Fz, F4, and F8. The temporal region contains T3, T4, T5, and T6. The central region contains C3, Cz, and C4. The parietal region contains P3, Pz, and P4. The occipital region contains O1 and O2. For each band, within-region and between-region wPLI means are computed. Compact graph features are then extracted from the region-level connectivity matrix, including mean node strength, clustering coefficient, global efficiency, and characteristic path length, which are commonly used to summarise EEG functional networks in dementia-related classification studies~\cite{si2023differentiating,ma2024classification,mlinari2026functional}. For graph construction, the weakest 70\% of edges are removed within each band and the remaining weighted graph is used for network descriptors. Graph descriptors are evaluated in the connectivity + graph ablation, while the selected multi-domain setting combines spectral, complexity, and pairwise connectivity descriptors.

\vspace{-3mm}
\subsection{Feature Selection and Dimensionality Control}
\vspace{-2mm}

Feature selection is performed inside each training fold to reduce high-dimensional EEG descriptors and mitigate overfitting in small-subject dementia datasets~\cite{akbar2025unlocking,rostamikia2024eeg}. No statistic estimated from test subjects is used for scaling, selection, or hyperparameter tuning. Features with variance below $10^{-6}$ are removed. Redundant features are removed when their absolute Spearman correlation exceeds 0.95, retaining the feature with the larger mutual-information score with the training labels. The remaining features are standardised by z-score using training-fold mean and standard deviation.

The primary selector ranks features by mutual information and selects $k \in \{50,100,200,400\}$ through inner GroupKFold validation. An $\ell_1$-regularised logistic selector is evaluated as a sparse embedded feature-selection ablation with $C_{\mathrm{reg}} \in \{0.01,0.1,1,10\}$. The final model always uses the selector chosen only from training subjects. This design controls feature dimensionality for spectral, complexity, and connectivity descriptors, which would otherwise exceed the number of independent subjects. Graph descriptors are evaluated separately in the connectivity + graph ablation.

\begin{table}[t]
\centering
\caption{Staged benchmark on the AD versus CN task. Stage A evaluates preprocessing, Stage B evaluates EEG segmentation and training-only augmentation, Stage C evaluates feature families, and Stage D evaluates dimensionality control and classifier choice. Dim. reports raw/selected feature dimensionality, such as 1596/100 for 1596 extracted descriptors reduced to 100 selected features. All metric values are percentages except Dim. and Time. MCC denotes Matthews correlation coefficient and is scaled by 100. Time is measured in seconds per subject. Rows marked by $\dagger$ are control settings used to separate the effect of training augmentation from the SVM regularisation grid.}
\vspace{-3mm}
\label{tab:staged_benchmark}
\scriptsize
\setlength{\tabcolsep}{2.5pt}
\resizebox{\textwidth}{!}{
\begin{tabular}{llrrrrrrrrr}
\toprule
Stage & Variant & Dim. & Acc. & BAC & F1 & Sens. & Spec. & AUC & MCC$\times$100 & Time \\
\midrule
\multicolumn{11}{l}{A. Preprocessing ablation with 10 s EEG epochs, spectral + complexity + connectivity features, MI top-100, and linear SVM} \\
Preprocess & Filtering only & 1596/100 & 78.46 & 78.21 & 80.56 & 80.56 & 75.86 & 84.21 & 56.42 & 22.8 \\
Preprocess & ICA-only & 1596/100 & 84.62 & 84.43 & 86.11 & 86.11 & 82.76 & 89.74 & 68.87 & 24.6 \\
Preprocess & ASR-only & 1596/100 & 83.08 & 82.71 & 84.93 & 86.11 & 79.31 & 88.41 & 65.67 & 25.1 \\
Preprocess & ASR + ICA & 1596/100 & 86.15 & 85.82 & 87.67 & 88.89 & 82.76 & 90.83 & 71.92 & 27.9 \\
\midrule
\multicolumn{11}{l}{B. EEG segmentation and augmentation with ASR + ICA, spectral + complexity + connectivity features, MI top-100, and linear SVM} \\
Segment & 5 s, no augmentation & 1596/100 & 83.08 & 82.71 & 84.93 & 86.11 & 79.31 & 88.96 & 65.67 & 31.5 \\
Segment & 10 s, no augmentation & 1596/100 & 86.15 & 85.82 & 87.67 & 88.89 & 82.76 & 90.83 & 71.92 & 27.9 \\
Segment & 20 s, no augmentation & 1596/100 & 84.62 & 84.43 & 86.11 & 86.11 & 82.76 & 89.61 & 68.87 & 23.7 \\
Segment & 30 s, no augmentation & 1596/100 & 81.54 & 81.32 & 83.33 & 83.33 & 79.31 & 86.92 & 62.64 & 20.4 \\
Aug. ctrl.$^\dagger$ & 10 s, no aug. + wider SVM $C_{\mathrm{reg}}$ grid & 1596/100 & 86.15 & 85.82 & 87.67 & 88.89 & 82.76 & 90.64 & 71.92 & 28.4 \\
Aug. ctrl.$^\dagger$ & 10 s, Gaussian noise only & 1596/100 & 86.15 & 85.82 & 87.67 & 88.89 & 82.76 & 90.95 & 71.92 & 28.8 \\
Aug. ctrl.$^\dagger$ & 10 s, amplitude scaling only & 1596/100 & 86.15 & 85.82 & 87.67 & 88.89 & 82.76 & 90.91 & 71.92 & 28.7 \\
Segment & 10 s, scaling + Gaussian noise & 1596/100 & 87.69 & 87.55 & 88.89 & 88.89 & 86.21 & 91.20 & 75.10 & 29.3 \\
\midrule
\multicolumn{11}{l}{C. Feature-family benchmark with ASR + ICA, 10 s EEG epochs, augmentation, MI selection, and linear SVM} \\
Feature & Spectral & 266/42 & 81.54 & 81.32 & 83.33 & 83.33 & 79.31 & 86.54 & 62.64 & 4.2 \\
Feature & Complexity & 475/35 & 76.92 & 76.82 & 78.87 & 77.78 & 75.86 & 81.37 & 53.49 & 9.6 \\
Feature & Connectivity & 855/46 & 80.00 & 79.93 & 81.69 & 80.56 & 79.31 & 85.92 & 59.69 & 13.4 \\
Feature & Connectivity + graph & 925/52 & 81.54 & 81.32 & 83.33 & 83.33 & 79.31 & 86.61 & 62.64 & 14.7 \\
Feature & Spectral + complexity & 741/73 & 84.62 & 84.43 & 86.11 & 86.11 & 82.76 & 89.37 & 68.87 & 13.9 \\
Feature & Spectral + complexity + connectivity & 1596/100 & 87.69 & 87.55 & 88.89 & 88.89 & 86.21 & 91.20 & 75.10 & 28.6 \\
\midrule
\multicolumn{11}{l}{D. Dimensionality control and classifier benchmark with ASR + ICA, 10 s EEG epochs, augmentation, and spectral + complexity + connectivity features} \\
Selector & No selection, linear SVM & 1596/1596 & 80.00 & 79.93 & 81.69 & 80.56 & 79.31 & 84.73 & 59.69 & 29.9 \\
Selector & MI top-50, linear SVM & 1596/50 & 84.62 & 84.43 & 86.11 & 86.11 & 82.76 & 89.02 & 68.87 & 27.5 \\
Selector & MI top-100, linear SVM & 1596/100 & 87.69 & 87.55 & 88.89 & 88.89 & 86.21 & 91.20 & 75.10 & 28.1 \\
Selector & MI top-200, linear SVM & 1596/200 & 86.15 & 85.82 & 87.67 & 88.89 & 82.76 & 90.12 & 71.92 & 30.0 \\
Selector & $\ell_1$ logistic, linear SVM & 1596/118 & 86.15 & 85.82 & 87.67 & 88.89 & 82.76 & 90.46 & 71.92 & 30.2 \\
Classifier & Shrinkage LDA & 1596/100 & 83.08 & 82.71 & 84.93 & 86.11 & 79.31 & 88.12 & 65.67 & 28.5 \\
Classifier & Logistic regression & 1596/100 & 84.62 & 84.43 & 86.11 & 86.11 & 82.76 & 89.55 & 68.87 & 30.4 \\
Classifier & Linear SVM & 1596/100 & 87.69 & 87.55 & 88.89 & 88.89 & 86.21 & 91.20 & 75.10 & 31.4 \\
Classifier & RBF-SVM & 1596/100 & 86.15 & 85.82 & 87.67 & 88.89 & 82.76 & 90.87 & 71.92 & 29.8 \\
Classifier & KNN & 1596/100 & 78.46 & 78.21 & 80.56 & 80.56 & 75.86 & 82.40 & 56.42 & 29.2 \\
Classifier & Random forest & 1596/100 & 84.62 & 84.43 & 86.11 & 86.11 & 82.76 & 89.06 & 68.87 & 29.9 \\
Classifier & ExtraTrees & 1596/100 & 86.15 & 85.82 & 87.67 & 88.89 & 82.76 & 89.94 & 71.92 & 28.3 \\
Classifier & Gradient boosting & 1596/100 & 83.08 & 82.71 & 84.93 & 86.11 & 79.31 & 87.63 & 65.67 & 27.7 \\
\bottomrule
\end{tabular}}
\vspace{-6mm}
\end{table}

\begin{table}[t]
\centering
\caption{Uncertainty analysis for key AD versus CN staged comparisons. Acc. and AUC values are percentages. The selected pipeline is ASR + ICA, 10 s EEG epochs with training-only augmentation, spectral + complexity + connectivity features, MI top-100 selection, and linear SVM.}
\vspace{-3mm}
\label{tab:uncertainty}
\scriptsize
\setlength{\tabcolsep}{2.5pt}
\resizebox{\textwidth}{!}{
\begin{tabular}{llrrrrr}
\toprule
Stage & Comparison setting & Acc. & Acc. 95\% CI & AUC & AUC 95\% CI & Paired $p$ \\
\midrule
Preprocess & Filtering only & 78.46 & [67.69, 87.69] & 84.21 & [74.38, 92.17] & 0.041 \\
Preprocess & ICA-only & 84.62 & [75.38, 92.31] & 89.74 & [82.31, 95.06] & 0.248 \\
Preprocess & ASR-only & 83.08 & [72.31, 90.77] & 88.41 & [80.65, 94.28] & 0.172 \\
Preprocess & ASR + ICA, no augmentation & 86.15 & [76.92, 93.85] & 90.83 & [83.46, 95.87] & 0.617 \\
\midrule
Feature & Spectral only & 81.54 & [70.77, 90.77] & 86.54 & [78.22, 93.05] & 0.109 \\
Feature & Complexity only & 76.92 & [66.15, 86.15] & 81.37 & [71.94, 89.65] & 0.018 \\
Feature & Connectivity only & 80.00 & [69.23, 89.23] & 85.92 & [77.40, 92.78] & 0.083 \\
Feature & Spectral + complexity & 84.62 & [75.38, 92.31] & 89.37 & [81.81, 94.95] & 0.271 \\
\midrule
Selection & No selection & 80.00 & [69.23, 89.23] & 84.73 & [75.98, 91.88] & 0.052 \\
Selection & MI top-50 & 84.62 & [75.38, 92.31] & 89.02 & [81.02, 94.62] & 0.289 \\
Selection & MI top-100 selected & 87.69 & [78.46, 95.38] & 91.20 & [84.01, 96.04] & Ref. \\
Selection & MI top-200 & 86.15 & [76.92, 93.85] & 90.12 & [82.14, 95.31] & 0.625 \\
Selection & $\ell_1$ logistic & 86.15 & [76.92, 93.85] & 90.46 & [82.67, 95.62] & 0.641 \\
\midrule
Classifier & Logistic regression & 84.62 & [75.38, 92.31] & 89.55 & [81.73, 94.84] & 0.251 \\
Classifier & RBF-SVM & 86.15 & [76.92, 93.85] & 90.87 & [83.26, 95.95] & 0.617 \\
Classifier & ExtraTrees & 86.15 & [76.92, 93.85] & 89.94 & [82.01, 95.19] & 0.508 \\
Classifier & Linear SVM selected & 87.69 & [78.46, 95.38] & 91.20 & [84.01, 96.04] & Ref. \\
\bottomrule
\end{tabular}}
\vspace{-6mm}
\end{table}

\vspace{-3mm}
\subsection{Classical ML Models}
\vspace{-2mm}

The classifier benchmark uses classical ML models because the dataset contains 88 independent subjects and the feature representation is designed for interpretability. The evaluated models are shrinkage linear discriminant analysis, logistic regression, linear support vector machine, radial-basis-function SVM, KNN, random forest, ExtraTrees, and gradient boosting.

Logistic regression uses class-balanced weights, $\ell_2$ regularisation, $C_{\mathrm{reg}} \in \{0.01,\\0.1,1,10\}$, and 5000 maximum iterations. Linear SVM uses class-balanced weights and $C_{\mathrm{reg}} \in \{0.01,0.1,1,10\}$. RBF-SVM uses $C_{\mathrm{reg}} \in \{0.1,1,10,100\}$ and $\gamma \in \{\mathrm{scale},0.001,0.01,0.1\}$. KNN uses $k \in \{3,5,7,11\}$ with distance weighting. Random forest and ExtraTrees use 500 trees, class-balanced weights, $\sqrt{p}$ feature sampling, and minimum leaf sizes in $\{1,2,4\}$. Gradient boosting uses $\{100,300\}$ estimators, learning rates in $\{0.05,0.1\}$, and maximum tree depths in $\{2,3\}$.

All classifiers are trained through a single pipeline containing scaling, feature selection, hyperparameter tuning, and classification. Models without calibrated probability outputs are wrapped by sigmoid calibration within the training fold. This allows the same subject-level probability aggregation rule to be used across all classifiers.

\begin{table}[t]
\centering
\caption{Fair LOSO baseline comparison, final task-level performance, and compact interpretation summary. The baseline rows are classical counterparts inspired by prior EEG dementia studies and are not reproduced results from the cited papers. All baseline and proposed rows use leakage-safe subject-level LOSO evaluation.}
\vspace{-3mm}
\label{tab:comparison_final}
\scriptsize
\setlength{\tabcolsep}{1.8pt}
\renewcommand{\arraystretch}{0.96}
\begin{tabular}{@{}L{0.095\linewidth}L{0.285\linewidth}c rrrr L{0.215\linewidth}@{}}
\toprule
Group & Method or task & Prot. & Acc. & BAC & F1 & AUC & Notes \\
\midrule
\multicolumn{8}{@{}l}{A. Classical counterparts under the same leakage-safe LOSO protocol} \\
Baseline & Band power + coherence + linear SVM~\cite{miltiadous2023dice} 
& LOSO & 83.08 & 82.71 & 84.93 & 88.30 & AD vs CN, 54 of 65 subjects \\
Baseline & Spectrum + complexity + synchronisation + RF~\cite{zheng2023diagnosis} 
& LOSO & 84.62 & 84.43 & 86.11 & 89.20 & AD vs CN, 55 of 65 subjects \\
Baseline & Rhythm-ratio discriminative features + SVM~\cite{rostamikia2024eeg} 
& LOSO & 74.58 & 73.97 & 78.08 & 79.40 & AD vs FTD, 44 of 59 subjects \\
Baseline & Electrode-pair communication features + RF~\cite{ma2024classification} 
& LOSO & 76.27 & 75.60 & 79.45 & 80.85 & AD vs FTD, 45 of 59 subjects \\
\midrule
\multicolumn{8}{@{}l}{B. Final performance of the selected benchmark pipeline} \\
Proposed & AD vs CN 
& LOSO & 87.69 & 87.55 & 88.89 & 91.20 & 57 of 65 subjects \\
Proposed & FTD vs CN 
& LOSO & 82.69 & 82.23 & 80.00 & 86.74 & 43 of 52 subjects \\
Proposed & AD vs FTD 
& LOSO & 77.97 & 77.23 & 81.69 & 82.31 & 46 of 59 subjects \\
Proposed & AD vs FTD vs CN 
& LOSO & 70.45 & 70.36 & 69.73 & 79.18 & 62 of 88 subjects \\
\midrule
\multicolumn{8}{@{}l}{C. Compactness and interpretation of the selected AD versus CN pipeline} \\
Feature & Spectral 
& LOSO & 81.54 & 81.32 & 83.33 & 86.54 & 42 features, theta-alpha, parietal \\
Feature & Complexity 
& LOSO & 76.92 & 76.82 & 78.87 & 81.37 & 35 features, MSE, temporal \\
Feature & Connectivity + graph 
& LOSO & 81.54 & 81.32 & 83.33 & 86.61 & 52 features, wPLI, frontal-temporal \\
Feature & Spectral + complexity + connectivity 
& LOSO & 87.69 & 87.55 & 88.89 & 91.20 & 100 features, temporal-parietal \\
\bottomrule
\end{tabular}
\renewcommand{\arraystretch}{1.125}
\vspace{-6mm}
\end{table}

\begin{figure}[t]
    \centering
    \includegraphics[width=\textwidth]{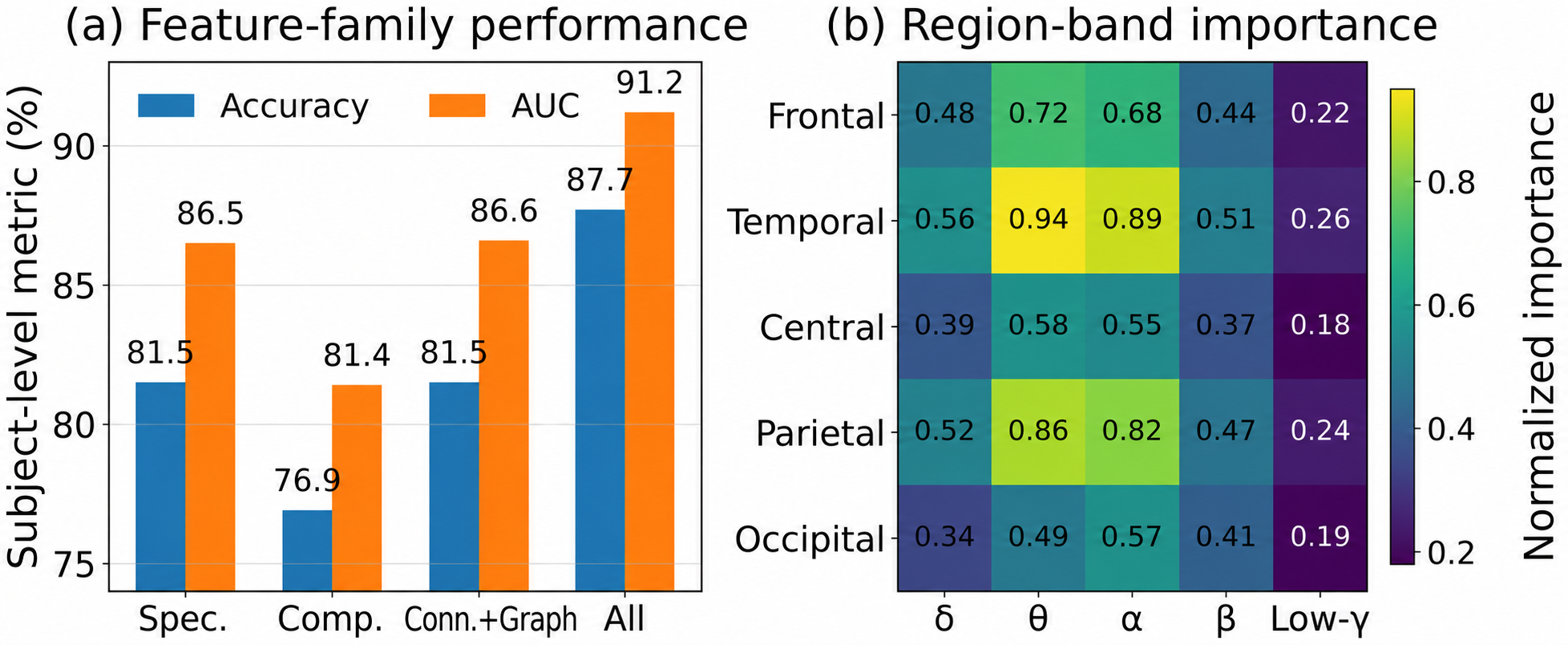}
    \vspace{-6mm}
    \caption{Visual summary of feature-family performance and region-band interpretation on the AD versus CN task. Panel (a) reports subject-level accuracy and AUC for selected features. Panel (b) reports normalised permutation importance aggregated by anatomical region and EEG frequency band using held-out subject-level predictions.}
    \label{fig:experiment_summary}
    \vspace{-6mm}
\end{figure}

%%%%%%%%%%%%%%%%%%%%%%%%%%%%%%%%%%%%%%%%%%%%%%%%%%%%%%%%%%%%%%%%%%%%%%%%%%%%%%%%%%%%%
\vspace{-3mm}
\section{Experiments}
\label{sec:experiments}
\vspace{-3mm}

This section evaluates the EEG-ML benchmark under the subject-independent protocol defined in Section~\ref{sec:Methodology}. The experiments report implementation settings, staged ablations over preprocessing, EEG segmentation, feature families, dimensionality control, and classifiers, followed by final task-level performance and interpretation of the selected pipeline.

\vspace{-3mm}
\subsection{Experimental Setup}
\vspace{-2mm}
Experiments were implemented in Python 3.10 using MNE-Python for EEG processing and scikit-learn for feature selection, calibration, and classification. All experiments were run on a workstation with 24 GB RAM. The outer evaluation used leave-one-subject-out (LOSO) cross-validation, and hyperparameter selection used an inner 5-fold GroupKFold over the training subjects. Within each outer fold, augmentation, scaling, feature selection, calibration, and classifier fitting were estimated from the training subjects only. 

Results are reported from subject-level predictions using accuracy, balanced accuracy, F1-score, sensitivity, specificity, Matthews correlation coefficient (MCC), and AUC. For binary tasks, the first class named in the task is treated as the positive class. For the three-class task, sensitivity, specificity, F1-score, and AUC are macro-averaged using a one-vs-rest scheme. Bootstrap confidence intervals and paired bootstrap tests are computed over held-out subject predictions. Runtime is measured as CPU feature-extraction and prediction time per subject.

\vspace{-3mm}
\subsection{Staged Ablation Study}
\vspace{-2mm}

The staged benchmark evaluates one pipeline component at a time while keeping the remaining settings fixed. Stage A compares preprocessing strategies, Stage B evaluates EEG-segment length and training-only augmentation, Stage C compares feature families, and Stage D evaluates dimensionality control and classifier choice. Unless a component is being tested, later stages use the best setting carried forward from the previous stage. Table~\ref{tab:staged_benchmark} reports all ablations on the AD versus CN task.

The best point estimate is obtained with ASR + ICA preprocessing and 10 s EEG epochs using training-only amplitude scaling and Gaussian-noise augmentation. This configuration combines spectral, complexity, and pairwise connectivity features with MI top-100 selection and linear SVM classification. The control rows show a one-subject gain over the no-augmentation setting, so the training-only augmentation is treated as a modest fold-internal regularisation effect. The uncertainty analysis in Table~\ref{tab:uncertainty} indicates that large feature-family gaps are stable, while small differences such as MI top-100 versus MI top-200 should not be overinterpreted. MI top-100 is therefore retained for its accuracy and compactness.

\vspace{-3mm}
\subsection{Fair LOSO Baseline Comparison}
\vspace{-2mm}

Table~\ref{tab:comparison_final} reports a fair baseline comparison designed to address validation-protocol mismatch. Instead of directly comparing our LOSO results with reported 10-fold CV results, the baseline rows present classical counterparts inspired by prior EEG dementia studies and evaluate them under the same leakage-safe LOSO protocol, subject-level aggregation rule, and training-fold feature-selection procedure used by the proposed pipeline. Under this protocol, the selected pipeline achieves the best AD versus CN result in the comparison, with 87.69\% accuracy and 91.20\% AUC. It improves accuracy by 4.61 percentage points over the band-power and coherence SVM counterpart and by 3.07 percentage points over the spectrum, complexity, and synchronisation RF counterpart.

\vspace{-3mm}
\subsection{Final Subject-Level Performance}
\vspace{-2mm}

Table~\ref{tab:comparison_final} also reports the final subject-level performance of the selected pipeline across all benchmark tasks. The highest performance is obtained on AD versus CN, with 87.69\% accuracy, 87.55\% balanced accuracy, 88.89\% F1-score, and 91.20\% AUC. Performance decreases on FTD versus CN and AD versus FTD, reaching 82.69\% and 77.97\% accuracy, respectively. The three-class task obtains 70.45\% accuracy, indicating that dementia-subtype and multi-class discrimination remain harder than disease-control classification under subject-level LOSO validation.

\vspace{-3mm}
\subsection{Best-Practice Recommendation and Interpretation}
\vspace{-2mm}

The final recommendation is to use the combined spectral, complexity, and connectivity feature set rather than a single feature family, followed by fold-internal mutual-information selection. Fig.~\ref{fig:experiment_summary} summarises the corresponding feature-family trend and region-band interpretation. The selected pipeline keeps 100 features from 1596 raw spectral, complexity, and pairwise connectivity descriptors and improves accuracy by 7.69 percentage points compared with using the unselected 1596-dimensional feature set. The interpretation indicates that spectral slowing features, complexity descriptors, and connectivity features provide complementary information, with the strongest region-level contributions concentrated in temporal-parietal and frontal-temporal regions.

\vspace{-3mm}
\section{Conclusion}
\label{sec:Conclusion}
\vspace{-3mm}

This paper presented a leakage-safe empirical benchmark for resting-state EEG-based dementia classification. The benchmark evaluates preprocessing, EEG segmentation, feature engineering, feature selection, classical classification, subject-level aggregation, and interpretation under subject-level validation. The best-performing pipeline combines ASR + ICA, 10 s EEG epochs with training-only amplitude scaling and Gaussian-noise augmentation, spectral, complexity, and pairwise connectivity features, mutual-information top-100 selection, and linear SVM, achieving 87.69\% accuracy and 91.20\% AUC on AD versus CN classification. Future work will validate the benchmark on external EEG cohorts and strengthen clinically grounded interpretation.

\vspace{-5mm}
\section*{Acknowledgements}
\vspace{-3mm}
This research was supported by the Curtin Health Innovation Accelerator Seed Funding Program.

\vspace{-3mm}
\bibliographystyle{splncs04}
\bibliography{main}

\end{document}